# 600-GHz Fourier Imaging Based on Heterodyne Detection at the 2nd Sub-harmonic

HUI YUAN,[1,*] ALVYDAS LISAUSKAS,[2,3] MARK D. THOMSON ,[1] AND HARTMUT G. ROSKOS[1,*]

[1]*Peer Review, Physikalische Institut, Goethe-Universität Frankfurt am Main, 60438, Frankfurt am Main, Germany*
[2]*Institute of Applied Electrodynamics and Telecommunications, Vilnius University, 10257 Vilnius, Lithuania*
[3]*Center for Terahertz Research and Applications (CENTERA), Institute of High Pressure Physics, Polish Academy of Sciences, 01-142 Warsaw, Poland*
[*]*yuan@physik.uni-frankfurt.de, roskos@physik.uni-frankfurt.de*

**Abstract:** Fourier imaging is an indirect imaging method which records the diffraction pattern of the object scene coherently in the focal plane of the imaging system and reconstructs the image using computational resources. The spatial resolution, which can be reached, depends on one hand on the wavelength of the radiation, but also on the capability to measure – in the focal plane – Fourier components with high spatial wave-vectors. This leads to a conflicting situation at THz frequencies, because choosing a shorter wavelength for better resolution usually comes at the cost of less radiation power, concomitant with a loss of dynamic range, which limits the detection of higher Fourier components. Here, aiming at maintaining a high dynamic range and limiting the system costs, we adopt heterodyne detection at the 2nd sub-harmonic, working with continuous-wave (CW) radiation for object illumination at 600 GHz and local-oscillator (LO) radiation at 300 GHz. The detector is a single-pixel broad-band Si CMOS TeraFET equipped with substrate lenses on both the front- and backside for separate in-coupling of the waves. The entire scene is illuminated by the object wave, and the Fourier spectrum is recorded by raster scanning of the single detector unit through the focal plane. With only 56 $\mu$W of power of the 600-GHz radiation, a dynamic range of 60 dB is reached, sufficient to detect the entire accessible Fourier space spectrum in the test measurements. A lateral spatial resolution of better than 0.5 mm, at the diffraction limit, is reached.

## 1. Introduction

Imaging with terahertz (THz) radiation is increasingly important for a wide range of scientific and industrial processes [1]. The unique nature of transparency of many non-polar substances is one of the reasons for the use of THz waves for nondestructive testing [2], others are the non-ionizing property and the spectral fingerprint of materials [3, 4]. Furthermore, THz imaging has a better spatial resolution [5] than microwave radar because of the shorter wavelength, thus is promising for short-distance imaging for robotics, autonomous vehicles and aerial exploration [6]. Recent improvements in the field of generation and sensing of THz radiation have led to advances in the technology of imaging, examples being compact on-chip imaging [7, 8], compressed sensing [9, 10], holography [5, 11–14], Fourier imaging [15, 16] and tomography [17]. The subject of the present publication is Fourier imaging. We briefly mention several of its characteristic features, which result from the Abbe theory of Fourier optics [18]. Note, that we will use the term *lens* for all kinds of imaging systems being employed (simple or composite transmission or reflection optics). First, Fourier imaging not only works in the real, but also in the virtual imaging regime of the converging lens. Second, it enables to record whole 2D and 3D scenes at the price of a reduced resolution even if only a limited number of pixels is recorded in the focal plane, which happens to represent – according to angular spectrum theory – the convolution of the propagated 2D spatial Fourier transformations of all object planes of the scene. Third, a typical feature of Fourier transforms is the rapid decline of the amplitude of the Fourier components for larger wave-vectors. If one desires to obtain a high spatial resolution of the image, one needs to record Fourier components at high wave-vectors, i.e., one must be able to reliably detect signals as far away from the optical axis of the lens as there are Fourier components available (Fourier components which have passed the wave-vector filtering by the lens [15]). This can only be achieved if the dynamic range of the detection is high.

Coherent detection is a requirement of Fourier imaging, but it also has drawbacks. The most obvious one is the increased cost and complexity of the imaging system, as it needs a powerful local-oscillator wave. If one strives for higher spatial resolution, utilizing short-wavelength radiation is the most direct way to achieve this. However, cost issues will then be more acute since the transmitters' prices increase with the frequency, whilst the output power drops off fast as the THz frequency rises. With the loss of beam power, also the contrast and the image quality deteriorate.

Sub-harmonic heterodyne detection (SHHD) can alleviate these challenges to a certain degree, as its LO beam uses radiation with a frequency which is a unit fraction – plus an offset – of that of the object wave. At the reduced frequency, it is easier and cheaper to obtain higher beam power. Here, we investigate high-dynamic-range SHHD for the first time with regard to Fourier imaging applications, building on the results of earlier research on sub-THz SHHD with AlGaAs/InGaAs/AlGaAs and CMOS FET detectors [19–22]. An aspect to keep in mind is that SHHD has an intrinsically lower conversion efficiency compared with heterodyne mixing at the fundamental frequency [20]. One hence has to ascertain that the higher available LO power is not offset by the loss in conversion efficiency. For a high sensitivity, it is important to guide all of the available radiation power onto the detector. When the object wave and the LO radiation are in-coupled from the same side onto the detector, one can easily loose a substantial amount of power by the beam combiner. In order to avoid such losses, we employ here a dual-substrate-lens approach for the beam coupling to the detector. The design consists of a composite wax/PTFE lens (super-hyperhemispherical shape [23]) on the front-side of the detector chip, and a conventional Si lens (hyperhemispherical shape) attached to the Si substrate. This dual-lens configuration allows the object wave and the LO beam to be coupled onto the detector from different sides, which in return not only avoids power loss, but also results in in a more compact detection unit since a beam combiner is avoided. With this innovation, the imaging system reaches a 60-dB dynamic range under object illumination with only 56-$\mu W$ of the 600-GHz radiation. The lateral spatial resolution is better than 0.5 mm, which already reaches the resolution limit at the given wavelength.

## 2. Experimental set-up

A schematic view of our experimental set-up is presented in Fig. 1(a). The radiation sources (S1 and S2) are multiplier chains. S1 works at 600 GHz (providing the object-illuminating radiation, vendor: RPG Radiometer Physics GmbH, base frequency of 16.66 GHz generated by a HP synthesizer, frequency multiplication factor: 36×, output power: 56 $\mu W$), and S2 at 300 GHz, frequency-locked to the first chain (serving as LO, vendor: Virginia Diodes, Inc., base frequency of 16.66 GHz + 1 kHz generated by a second HP synthesizer, frequency multiplication factor: 18×, output power: 600 $\mu W$). Prior to the illumination of the target, the 600-GHz radiation is collimated with a plane-convex Teflon lens (L1, 10-cm focal length and 4-inch aperture). The resultant beam diameter is about 4 cm. Another Teflon lens (L2, 6.5-cm focal length and 4-in. aperture) focuses the transmitted radiation onto the backside of the chip with the TeraFET detector, which is raster-scanned in the focal plane of L2. The TeraFET consists of a CMOS field-effect transistor integrated with a broad-band bow-tie antenna which permits it to cover both the imaging and LO frequencies [24]. The radiation from the 300-GHz source is focused with a third Teflon lens (L3, 2.5-cm focal length, 2-in. aperture) onto the front-side of the active region of the detector chip through a wax/PTFE superstrate lens. The TeraFET acts as a receiver [25, 26]; in sub-harmonic detection mode, it provides the difference-frequency signal of the radation from S1 and that of the 2$^{nd}$ harmonic of the radiation from S2. The signal is then processed with a lock-in amplifier (PerkinElmer, integration time: 20 ms, providing a dynamic range of 60 dB) whose reference signal at 1 kHz is obtained by mixing the drive signals from the two synthesizers in a MARKI electrical mixer. The lock-in amplifier works at the 36$^{th}$ harmonic of the reference signal. Unlike [15], we do not attenuate the beam in order to suppress standing waves; their effect is present in the images.

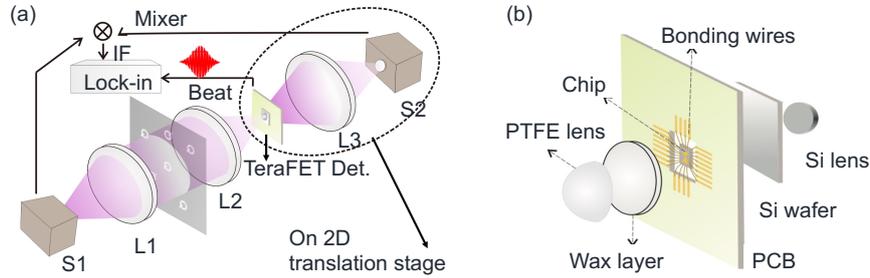

Fig. 1. (a) Schematic of the experimental set-up for transmission-mode Fourier imaging. (b) Exploded view of the detector unit. The object wave impinges onto the TeraFET from the backside through a Si substrate lens (right side in this figure), while the LO wave is front-side-coupled through the wax/PTFE lens.

A 3D exploded drawing of the detector unit is shown Fig. 1(b). The TeraFET chip is pasted onto a highly resistive Si substrate which is then attached to a PCB board with a cut-out as shown in the drawing. The PCB board provides the bonding pads for the external wiring of the TeraFET. After wire bonding of the TeraFET, a hemispherical Si lens with a 4-mm diameter is aligned in the THz beam and glued onto the Si substrate (it should be noted that we obtain a better responsivity of the detector with larger substrate lenses, but the smaller lens allows for the small pixel size desired for the recording of the Fourier space spectrum during imaging). Subsequently, the composite wax/PTFE lens is integrated on the front-side of the detector chip. A detailed description of the design and fabrication of that novel type of substrate lens will be published elsewhere [27]. In short, a 2-mm-thick paraffin wax layer is dropcast onto the TeraFET by dripping molten wax onto it, allowing to embed the bonding wires within the wax material. The wax, after hardening, forms a protection layer, onto which a custom-made PTFE hemispherical lens (diameter: 8 mm, pedestal height: 2 mm, leading to a vertical center thickness of 6 mm) is attached. The paraffin wax and the PTFE material have similar refractive indices at THz frequencies which minimizes reflection losses at the interface. Care has to be taken to correctly center the lens on the detector. This is done during active operation of the detector, maximizing the rectified signal by alignment of the PTFE lens. The lens is then fixed by careful re-melting and curing of the surface of the paraffin wax layer. The surface melting is achieved with the help of a heat gun, whose warm air is directed towards the PTFE lens which transfers the thermal energy via heat conduction to the wax.

The detector unit together with lens L3 and radiation source S2 are mounted on a translation stage, which allows us to perform raster-scan recording of the amplitude and phase of the radiation field across the focal plane of lens L2. In the set-up as described above, we cover a scan area of $60 \times 60$ mm$^2$. Using 1-mm scan steps in both lateral directions, the total time of data recording is 15 minutes (single frame). The useful size of the detection area, which determines the image resolution, in practice depends on the noise-equivalent power (NEP) of the detector and the dynamic range reached with the imaging set-up. From the data, the image is reconstructed numerically by applying the MATLAB program for inverse Fast Fourier Transformation.

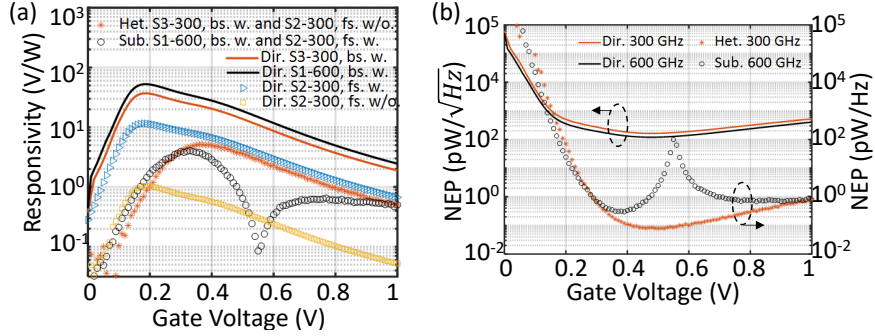

Fig. 2. Measured TeraFET performance data as a function of the FET channel's gate bias voltage: (a) Responsivity $\Re$ of the TeraFET, (b) noise-equivalent power (NEP). In the legends, "Dir." stands for direct power detection, "Het." for fundamental heterodyne detection, and "Sub." for $2^{nd}$-order sub-harmonic heterodyne detection (SHHD). THz radiation is either coupled onto the detector from the backside of the chip ("bs.") with a 4-mm diameter Si lens or from the front-side ("fs.") both with and without a wax-PTFE superstrate lens being attached to the chip. The RF and LO beams and their respective radiation frequencies are indicated by the radiation source used and the frequency: "S1-600", "S2-300" and "S3-300".

## 3. Results

### 3.1. Sub-harmonic Heterodyne Detection

In order to assess the performance of SHHD with the dual-lens approach, Fig. 2 presents detector sensitivity data recorded in different modes of operation. The results are shown as a function of the FET's gate bias voltage, revealing also the different optimal operation points.

Fig. 2(a) presents the responsivity of the TeraFET for six different measurement scenarios. Four of the curves of Fig. 2(a) display results obtained by power detection (also called direct detection, marked in the figure as "Dir."), and another two show data of heterodyne measurements (marked in the figure as "Het." and "Sub.", respectively). We first address power detection at 300 GHz (red-brown full line) and 600 GHz (black full line) with the radiation impinging from the detector chip's backside through the Si substrate lens (the wax/PTFE superstrate lens was attached, but not used, during these measurements). For the 300-GHz measurements, we employ 56 $\mu$W of radiation from a second 300-GHz source, S3, which is from RPG Radiometer Physics GmbH (base frequency of 16.66 GHz generated by a HP synthesizer, frequency multiplication factor: 18×). The responsivity is found to be very similar at the two radiation frequencies, which testifies for both the smooth frequency dependence of the detector design, already reported about in [24], and the good alignment of the Si substrate lens. We now turn to power detection at 300 GHz for the case where the radiation impinges onto the detector's front-side. For these measurements, we used radiation source S2 at 600 $\mu$W. The ocher-colored squares (respectively blue triangles) in Fig. 2(a) present measurements taken without (with) a wax/PTFE superstrate lens attached to the detector. The lens improves the responsivity by more than one order of magnitude, from 0.419 V/W without wax/PTFE lens to 5.063 V/W with lens, these data taken at the gate voltage of 0.48 V of the best NEP value. However, compared to backside coupling (red-brown line, responsity value: 15.02 V/W at 0.48 V), the responsivity remains weaker by about a factor of three. This can have several reasons, for example higher reflection losses due to refractive-index mismatch of the paraffin wax ($n$=1.52) and of PTFE ($n$=1.42) compared with the Si substrate ($n$=3.4), absorption losses in the wax, slight misalignment of the lens stack, etc. [27].

The final two curves of Fig. 2(a) are recorded in heterodyne mode. The red-brown full dots represent fundamental heterodyne detection of 300-GHz radiation, the open black circles 2$^{nd}$-order SHHD of 600-GHz waves. For the former, we employ source S3 to provide the RF radiation (power: 56 $\mu$W). For the LO wave, 600 $\mu$W of 300-GHz radiation is used both in the case of fundamental heterodyne detection at 300 GHz as well as for SHHD at 600 GHz. For the latter (the SHHD measurements of Fig. 2(a)), we apply a wax/PTFE superstrate lens, but do not so in the case of the former. There, we rather reproduce our Fourier imaging situation of [15], where the object-illumination wave at 300 GHz impinged through a Si substrate lens onto the backside of the TeraFET, whereas the 300-GHz LO radiation was in-coupled from the front without a superstrate lens. We find that the different measurement conditions lead to about the same peak responsivity of both measurements modalities (see below), respectively dynamic range (cf. Sec. 3.5), and that this dynamic range is sufficient to achieve a near-diffraction-limited spatial resolution with Fourier imaging (cf. Sec. 3.2).

Comparing the responsivity curves of the two cases, one observes different gate voltage dependencies. For heterodyne detection at the fundamental frequency, the dependence is similar to that of power detection, however with the peak responsivity shifted to higher gate voltage (0.3-0.4 V instead of 0.1-0.2 V). For SHHD, the dependence is quite different. After a peak at about 0.3 V, the responsivity drops to zero and then reappears with reversed sign (sign reversal not shown in the plot, which displays the absolute value of the responsivity). The sign change results from the gate voltage dependence of the channel impedance [20]. The data displayed in the figure show that we achieve quite similar maximal responsivity values for the two cases. And this despite the intrinsically lower conversion efficiency of SHHD versus heterodyne detection at the fundamental [20]. One can conclude that the lower responsivity which one expects for SHHD in comparison with heterodyne detection at the fundamental for the same amount of LO radiation, is nearly compensated by the tighter focusing of the LO radiation with the wax/PTFE superstrate lens.

The NEP as a function of the gate voltage is calculated from the DC resistance $R$ of the detector and the responsivity $\Re$ (Fig. 2(a)) with the expression NEP$_d = \sqrt{4k_BTR}/\Re$ (for power detection) and NEP$_h = 4k_BTR/\Re^2$ (for fundamental heterodyne detection and SHHD, derived on the basis of [28]), where $k_B$ is the Boltzmann constant and $T$ stands for room temperature. The resulting NEP curves are shown in Fig. 2(b). For power detection at 600 GHz, the best NEP$_d$ value of 120 pW/$\sqrt{Hz}$ is found at a gate voltage of 0.48 V. At 300 GHz, the corresponding value is 163 pW/$\sqrt{Hz}$, at the same gate voltage. As mentioned before, the performance does not reach the optimal characteristics which we achieve with larger-size Si lenses, but the smaller 4-mm Si lens is necessary for the imaging application in this work to guarantee the small pixel size needed for a good image quality. For comparison, one reaches an NEP$_d$ of 48 pW/$\sqrt{Hz}$ with radiation at 600 GHz and a well-aligned Si substrate lens with a diameter of 12 mm [24, 29]. The best NEP$_h$ values are 0.08 pW/Hz for fundamental heterodyne detection (gate voltage: 0.50 V) and 0.3 pW/Hz for SHHD (gate voltage: 0.40 V), respectively. It is obvious that fundamental heterodyne detection and SHHD have at least a 33-dB, respectively 26-dB better NEP for a 1-Hz bandwidth compared with power detection. Care has to be taken, that the optimal gate-voltage operation points of the different detection modes are different from each other.

Fig. 3 displays the detector response in units of mV as a function of the radiation power for three modes of operation of the TeraFET. We compare power detection (red-brown dots) and SHHD (blue dots) at 600 GHz with heterodyne detection at the fundamental at 300 GHz (purple dots). In all cases, the signal wave is backside-coupled to the detector through a Si substrate lens (diameter: 4 mm), while the LO radiation impinges from the front-side through a wax/PTFE superstrate lens (diameter: 8 mm). Note that this is the only instance where we apply a superstrate lens for fundamental heterodyne detection at 300 GHz. In each case, the data are obtained by gradual attenuation of the power of the (object-illumination, "RF") radiation, the maximum power being 56 $\mu$W for both the 600-GHz and the 300-GHz wave. For both SHHD and fundamental heterodyne detection, a fixed LO power of 600 $\mu$W at 300 GHz is employed. No filters for standing-wave suppression are used in these measurements. For each mode of operation, the respective optimal gate bias point of the TeraFET is chosen. The data reveal fundamental heterodyne detection to be superior to SHHD, as expected from the discussion of Fig. 2. Direct (power) detection is superior at the highest beam power, but rolls off linearly with decreasing power, while the two heterodyne modes exhibit a flatter square-root dependence on power. The cross-over points are at -15 dBm for heterodyning at the fundamental, and at -25 dBm for SHHD. Given the noise floor of -50 nV, which is determined here by the noise properties of the lock-in amplifier and not the detector unit (see, in contrast [30]) the dynamic range of $2^{nd}$-order SHHD amounts to 60 dB, which is by 20 dB larger than the dynamic range achieved with power detection. This performance provides a satisfactory basis for our Fourier imaging experiments, for which a sufficiently large dynamic range is required in order to detect the weak signals at the high-wave-vector Fourier components. The required dynamic range is discussed and quantified in the next section.

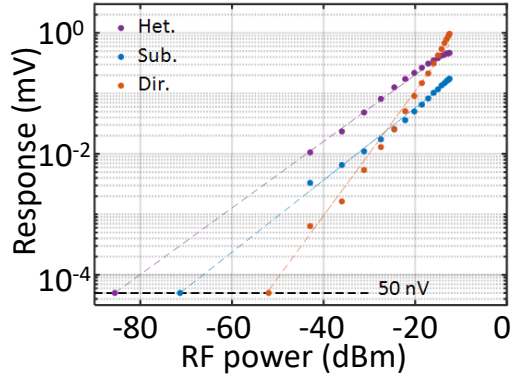

Fig. 3. TeraFET response to 600-GHz and 300-GHz radiation as a function of the radiation power. In the legend, "Dir." stands for power detection of 600-GHz radiation, "Sub." for $2^{nd}$-order sub-harmonic heterodyne detection (SHHD) of 600-GHz radiation, and "Het." for fundamental heterodyne detection of 300-GHz radiation.

### 3.2. Dynamic Range and Image Quality

The relationship between the dynamic range of the detection process and the quality of the image calculated directly (without advanced image processing) from the measured Fourier space data is elucidated by the following experiment. We record the Fourier space spectrum of the object with our maximal available dynamic range, and then filter out those Fourier data whose magnitude is below a chosen threshold value. In order to take into account that low-dynamic-range data are also more noisy in the useful spectral range, equivalent noise is added to the unfiltered data. The noise is generated as random complex values of both the amplitude and phase, with a magnitude threshold of -40 dB or -20 dB (corresponding to the dynamic range) of the maximum magnitude of the data. With this numerical magnitude filter and noise inclusion, we simulate measurements performed at various values of the system's dynamic range.

The object of this study is the planar metallic grid shown in Fig. 4(a). The grid has a period of 7.5 mm and a stripe width of 2.5 mm. The grid is placed perpendicular to the beam path, 6 cm in front of L2. The Fourier space spectrum is recorded with the measurement system's full dynamic range of 60 dB. The resultant intensity and phase maps are shown in Fig. 4(b) and (e), respectively. One readily discerns useful signal levels across the entire scanned focal-plane area. The calculated intensity and phase images shown in Fig. 4(h) and (k) reconstruct the object well over the entire cross-sectional area of the THz beam (diameter: 4 cm).

In a next step, the thresholding operation is carried out. For pixels with an intensity less than a threshold value of 40 dB, the pixel intensity value is replaced by the threshold value, and the phase is given a random value. The resultant filtered Fourier data are shown in Fig. 4(c) and (f), respectively. The number of useful pixels has diminished considerably. The reconstructed intensity and phase images in Fig. 4(i) and (l) still yield a proper rendition of the object close to the optical axis, but the image quality has deteriorated at the edges of the beam's cross-section. Selecting an even lower threshold of 20 dB leads to the Fourier space pseudo-spectra of Fig. 4(d) and (g), and the reconstructed intensity and phase images of Fig. 4(j) and (m). The number of useful pixels in the Fourier plane has become small. Concomitantly, the object rendition is successful only very close to the optical axis, but also there, the contrast is diminished and the spatial resolution degraded.

These results confirm and illustrate the statement made above: With a higher dynamic range of the measurement system, more pixels can be recorded reliably in the Fourier space. This has mainly two consequences. First, a high dynamic range is required for a proper reconstruction of the scene across the entire illuminated real-space area. If the system's dynamic range is low, the off-axis regions of the illuminated real space, where the THz beam is weak, blur or even vanish in the reconstructed image. Second, a high dynamic range will enable pixels representing large wave-vectors to contribute to the image reconstruction, although their intensity tends to be low. They are important for the achievement of a high lateral spatial resolution.

The loss of contrast and spatial resolution with decreasing dynamic range is highlighted in the line scans of Fig. 4(n) and (o) showing horizontal one-dimensional intensity and phase plots extracted from the reconstructed grid images of Fig. 4(h), (i), (j), (k), (l) and (m), respectively. While the high-dynamic-range data (blue lines) exhibit well-resolved image structures with a contrast up to 35 dB, the maximal contrast decreases to 25 dB in the case of the reduced 40-dB dynamic range (red-brown lines). The grid structure is still resolvable, the spatial resolution has not degraded significantly yet. For the lowest dynamic range of 20 dB (black dashed lines), only the structures in the center area are barely resolvable, the resolution is poor, and the maximal contrast is now about 15 dB. Here – and even more so in the Fourier spectra –, one can also see that the phase tends to be more sensitive to the noise than the intensity.

An important conclusion which one can draw from the information of this section is, that the gain in dynamic range [15], obtained by the dual-lens approach for the illumination of the TeraFET detector, is very valuable with regard to the achievable image quality. The 60-dB dynamic range of the imaging system should allow in many cases to record and usefully exploit the full Fourier spectrum in the focal plane of the imaging lens. Judging from the reconstruction achieved with the 40-dB-range Fourier spectral data of Fig. 4, even a certain amount of beam attenuation, e.g. by absorption or scattering, can be tolerated.

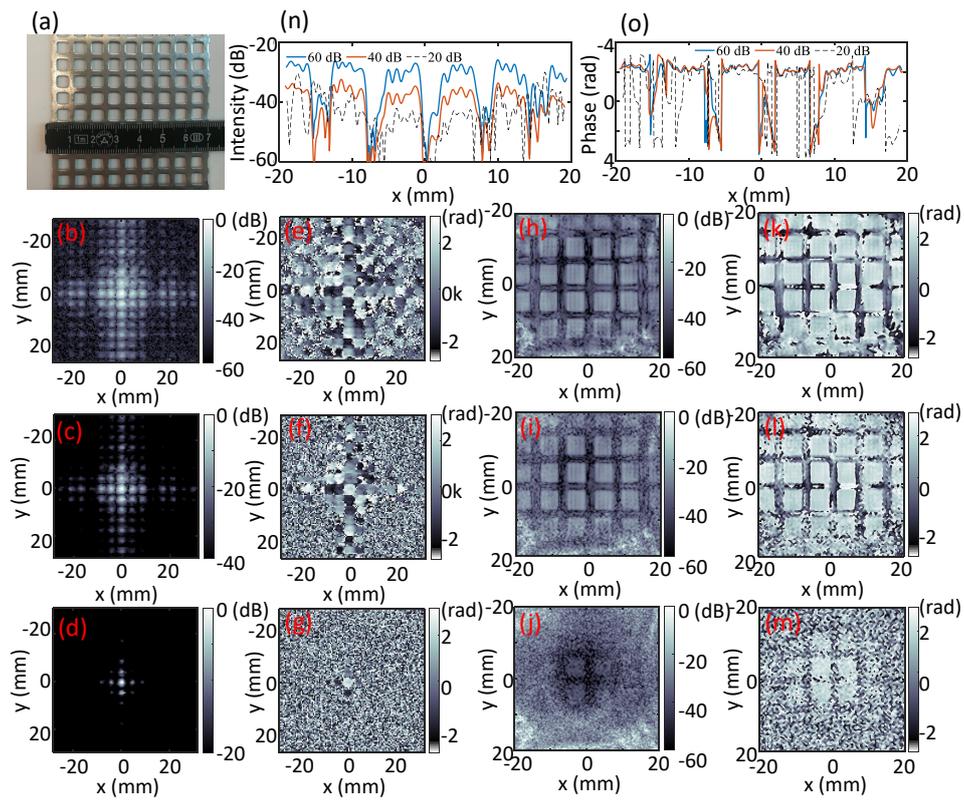

Fig. 4. Photographic image, measured Fourier spectral data and image reconstructions of a metal grid: (a) photograph of the grid; (b)–(d) intensity Fourier spectra; (e)–(g) phase data of the Fourier spectra; (h)–(j) intensity plots of the image reconstructions; (k)–(m) phase image reconstructions. (n) and (o) show one-dimensional line scans obtained by a horizontal pass through the middle of the reconstructed intensity and phase images, respectively. In the legends, "60 dB", "40 dB" and "20 dB" stand for the degree of thresholding of the Fourier data explained in the text.

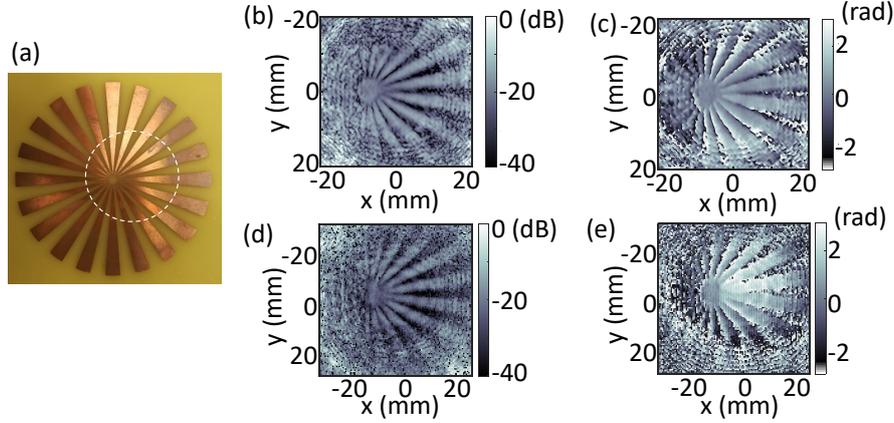

Fig. 5. Photograph of the Siemens star, reconstructed Fourier images thereof and plane-to-plane images: (a) Photograph of the Siemens star fabricated from the copper metal of a printed circuit board; (b) reconstructed intensity and (c) phase image based on Fourier spectrum recording in the focal plane of the imaging lens; (d) intensity and (e) phase image recorded in the image plane. The images of (d) and (e) are re-scaled to have the same size as the Fourier images of (b) and (c).

### 3.3. Fourier Imaging Compared with Plane-to-plane Imaging

In this section, we compare computational Fourier imaging with conventional plane-to-plane imaging (i.e., recording of the data in the image plane) based on the example of a planar metal structure as imaged object. In both cases, amplitude and phase data are recorded with the sub-harmonic 600-GHz/300-GHz detector unit.

A Siemens star is used as the specimen for the comparative measurements. The object is displayed in Fig. 5(a). The star is deliberately placed off-centric into the THz beam in order to cover a larger range of widths of the metal triangles which make up the start shape. The 3600-pixel Fourier data (1-mm pixel pitch) and the 14400-pixel plane-to-plane image (0.465-mm pixel pitch) are recorded each in a single raster-scan trace with a lock-in time constant of 20 ms. In both cases, the Siemens star is placed at an object distance of 10 cm in front of the imaging lens L2 (cp. Fig. 1). Fourier imaging records the data in the focal plane at the focal distance of 6.5 cm of L2. From the 3600-pixel Fourier spectrum, we calculate a 14400-pixel image (applying zero padding in the Fourier transformation for the expansion of the pixel number). For plane-to-plane imaging, the detector is placed into the image plane which is 18.6 cm away from L2 for the given object distance. This yields a magnification factor of 1.86 in each lateral direction. The image is then re-scaled to the size of the reconstructed Fourier image. The resultant intensity and phase images of the Siemens star are displayed in Fig. 5(b)–(e). All images are corrected for the cross-sectional intensity and phase profile of the illumination beam (for further details, see below).

We want to point out three salient aspects of Fig. 5. (i) **Dynamic range**: Both image-recording modalities yield images with the same dynamic range of 40 dB. Fourier imaging is not inferior to plane-to-plane imaging. (ii) **Sparsity:** An interesting feature of Fourier imaging is that one can generate images with variable numbers of pixels in the reconstruction process. Here, we have recorded *only a quarter of the total number of pixels of plane-to-plane imaging*, but have reconstructed an image with the same number of pixels at hardly discernible loss of imaging quality. In this process, we benefit from the sparse nature of the data in Fourier space [9, 31]. It is expected that the pixel number could be further compressed in Fourier space without significant loss of information. This is a subject for further studies. (iii) **Aberrations affect the images differently:** The focusing lens L2 is a planar-hyperbolic PTFE lens, and it is placed in such a way into the beam path that the collimated THz beam impinges onto the side with the planar surface. The advantage of using L2 over a lens with spherical surfaces is that it is thinner (hence has less absorption) for the same small focal length. While such a lens focusses a collimated beam, which propagates parallel to the optical axis of the lens, perfectly, aberrations occur and the effective focal length changes, if the angle between the beam path and the optical axis of the lens is changed. As a result, the focal area is not planar, but curved. Combined with the effect of multiple reflections at the interfaces of L2 and other optical components in the beam path, this leads to the circular ring pattern (with the origin on the optical axis of the lens) which can be observed in all four pictures of Fig. 5(b)–(e). We note, that – in the case of Fourier imaging – the effect of the curved focal area can in principle be accounted for numerically after mapping out of the focal area. This will not be pursued further here. While the curved focal area of L2 affects the two imaging modalities in a similar way, the sum of the THz beam deformations, which also may be incurred by optical components in the beam path other than L2, influence the final images differently. The reason lies in the different post-processing of the raw data for the correction of the intensity and phase profile of the illumination beam (note, that all images in this paper have been processed in the way to be described now). In the case of plane-to-plane imaging, the correction proceeds in the following way. First, the intensity and phase distribution of the bare illumination beam (no object in the beam path) is measured in the imaging plane. Then, the correction of the object image is performed by dividing the intensity distribution of the object image by that of the bare-beam image, and by subtracting the phase of the latter from that of the object image. In the case of Fourier imaging, the procedure is different. There, the field amplitude and phase maps of the bare illumination beam are measured in the focal plane of L2, yielding the point-spread function (PSF) of the imaging system for an on-axis object point with an infinite object distance. This PSF is then used in a deconvolution algorithm with the Fourier spectrum of the object image in the Fourier domain, prior to the reconstruction of the image by back-transformation. Comparing the resultant images shown in Fig. 5, one notices two differences. First, the intensity plot of the image obtained by Fourier imaging shows more detail and sharper features in the center region of the Siemens star than the corresponding plane-to-plane image (cp. Fig. 5(b) and (d)). The difference is also mainfest in the phase plots (cp. Fig. 5(c) and (e)). We explain this observation by a better correction of aberrations achieved by the PSF-based deconvolution process of Fourier imaging as outlined above. In the phase plots, one also finds that the features of the Siemens star are better reproduced in the outer regions of the image. This may point to a better correction for coma by the PSF-based deconvolution.

*3.4. Lateral Resolution*

To validate the resolution performance of the 600-GHz SHHD Fourier imaging system, four segments of a 1951 USAF resolution test pattern (realized as a metal structure on a printed circuit board) are adopted as the imaging object. Photographs of the object – with indicators of the imaged areas on the test pattern – and the reconstructed images are displayed in Fig. 6. The photographs in Fig. 6(a) clearly show the six paired stripe elements of Group -2 (left side of the left plot of Fig. 6(a)) and Group -1 (right side of the left plot of Fig. 6(a)), more difficult to discern are the smaller elements of Group 0 and Group 1 located between these two groups. Intensity images are shown in Fig. 6(b)–(e), phase maps in Fig. 6(f)–(i). The dynamic range is found to amount to 30 dB. Fig. 6(d) and (h) and Fig. 6(e) and (i) investigate only the larger patterns of Group -1, whereas Fig. 6(b) and (f) and Fig. 6(c) and (g) include the smaller features of Group 0 and Group 1. Metal stripes in lateral and vertical directions are resolvable down to the patterns in Group 0 Element 1 with a stripe and interval width of 0.5 mm. Stripes of Group 0 Element 2 and 3 (stripe and interval width of 0.445 mm and 0.397 mm, respectively) are ambiguous, although they would probably be resolvable with the prior knowledge of the stripe pattern by using artificial intelligence (AI) technology [32, 33].

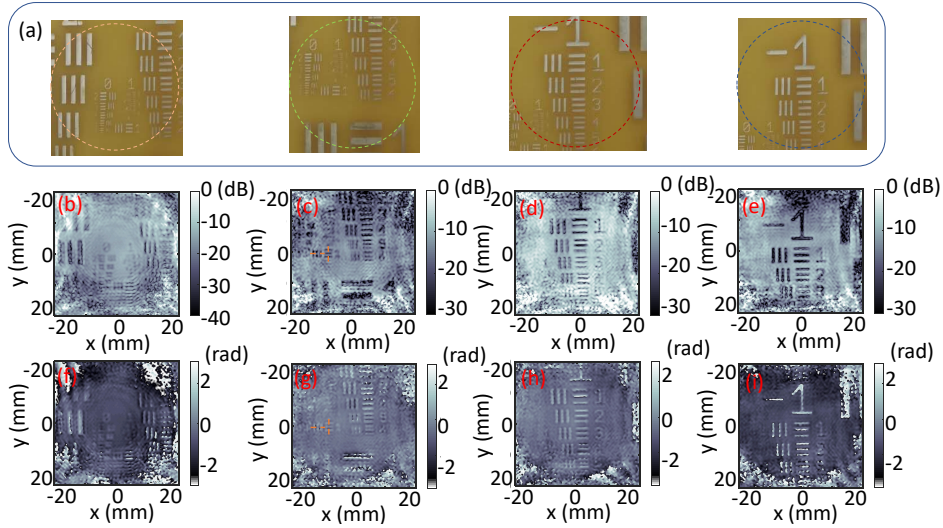

Fig. 6. Photograph and reconstructed images of a 1951 USAF resolution test chart: (a) photograph of the metal pattern with the imaged regions marked by circles plotted by dashed lines; (b), (c), (d) and (e) reconstructed intensity images of the four different regions of the chart indicated in (a); (f), (g), (h) and (i) reconstructed phase images relevant to (b), (c), (d) and (e) respectively.

Fig. 7 displays horizontal and vertical line scans through Element 1 of Group 0 in the images of Fig. 6(c) and (g) as indicated by the orange dashed lines. Fig. 7(a) shows horizontal cuts through the vertical stripes (blue line: intensity plot, red-brown line: phase plot). Fig. 7(b) corresponding vertical cuts through the horizontal stripes. The stripes manifest themselves as oscillations, which indicates that the spatial resolution is better than 0.5 mm. This value is equal to the wavelength of 600-GHz waves and is close to the resolution of 0.4 mm expected from the Rayleigh criterion for the given lens diameter (10 cm) and focal length (6.5 cm) [34]. The intensity and the contrast for the vertically oriented stripes are larger than those for the horizontally oriented ones. This is a result of the vertical linear polarization of the 600-GHz radiation and is explained by diffraction. Electromagnetic waves experience strong diffraction when they interact with periodic metal structures where the stripes are oriented perpendicular to the polarization of the radiation and where the period is similar to the wavelength. In this case, the signal transmitted in forward direction is weak. Radiation polarized parallel to the stripes correspondingly is expected to produce a stronger forward-transmitted signal. This observation suggests that the Fourier imaging system is also suitable for polarization imaging, which may open potential for applications in the industrial field.

### 3.5. Comparison with Fourier Imaging at 300 GHz

We now compare the spatial resolution obtained with the 600-GHz SHHD Fourier imaging system with that of the 300-GHz fundamental-heterodyne system. The focal length of L2 is different in the two cases. The 600-GHz system uses a L2 lens with a focal length of 6.5 cm, the 300-GHz system one with 15 cm. The beam diameter is 4 cm for the 600-GHz system and 5.5 cm for the 300-GHz system. In the case of the 600-GHz system, the Fourier spectrum is recorded over an area of 60×60 mm$^2$, and over an area of 80×80 mm$^2$ with the 300-GHz system, in each case with a pixel pitch of 1×1 mm$^2$. The 300-GHz imaging system uses an illumination beam with a power of 1 mW, but attenuated by 60% (in order to suppress standing waves) [15]. The LO beam with a power of 432-$\mu$W is coupled to the front-side of the TeraFET without the use of a superstrate lens.

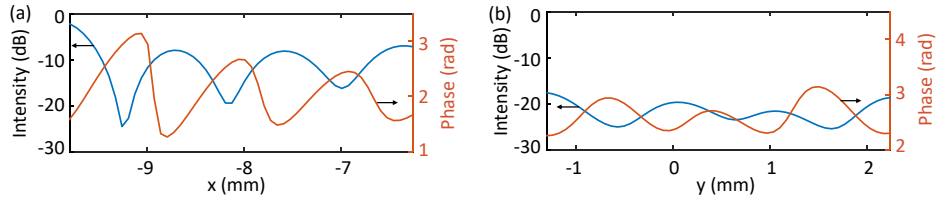

Fig. 7. One-dimensional data extracted from Fig. 6(c) and (g), see orange dashed lines there. Cuts are perpendicular to the vertical stripes (a) and the horizontal stripes (b), respectively, of the patterns in Group 0 Element 1 of the USAF 1951 resolution pattern. Intensity plots are displayed in blue color, phase plots in red-brown.

The comparison is performed on the basis of images taken of the same region of the USAF resolution test chart (with Elements 1 to 6 of Group -1 at the top and the pattern of Element 1 of Group -2 at the bottom). The Fourier spectra and the reconstructed images are plotted together in Fig. 8. The top row displays result for illumination at 600 GHz, the bottom row those for 300 GHz. The two left images of each column are the intensity and phase maps recorded in the focal plane. The right two images are the reconstructed intensity and phase images of the object. The images for 600 GHz (Fig. 8(c) and (d)) are the same ones already shown in Fig. 6(c) and (g).

The plots in Fig. 8(g) and (h) show that, at 300 GHz, only the pattern of Element 1 of Group -2 is resolved, while the images of the elements of Group -1 are blurred. As the stripe width of Element 1 of Group -2 is 2 mm, the resolution can be specified as being better than 2 mm (cp. resolution of 1.875 mm according to the Rayleigh criterion) [15]. The spatial resolution is hence close to the diffraction limited value.

We also point out again the advantage arising from the additional wax/PTFE superstrate lens for the 600-GHz system. The improved in-coupling of the LO radiation to the detector has the consequence that the 600-GHz system with its 56-$\mu$W of illumination power exhibits a comparable dynamic range as the 300-GHz system with its much larger illumination power of 1 mW (which, however, is attenuated in that system by 60% for suppression of standing wave, which is not done in the 600-GHz system). However, it should be noted that there is also a drawback associated with the introduction of the superstrate lens, as it leads to the appearance of standing waves in the detector unit and the imaging system. On the detector level, the curved surfaces of the wax/PTFE superstrate lens and the Si substrate lens form a cavity in which standing waves develop. This is the case for both the illumination radiation and the LO wave. With respect to the entire imaging system, the superstrate/substrate lens pair extends the transmission path for both the illuminating and LO radiation. As a consequence, each beam experiences more reflecting elements in its effective beam path, which promotes the development of standing waves in the system. A possible way to alleviate this problem is to introduce an optical isolation device with low insertion loss, another the shortening of the coherence length of the radiation.

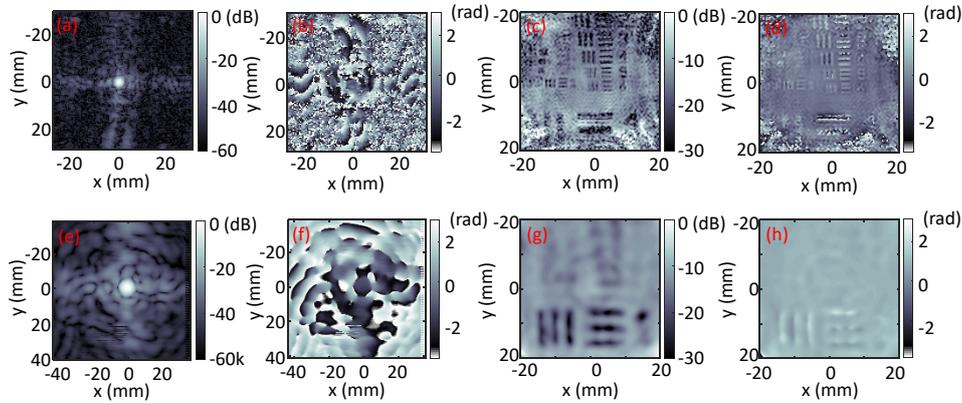

Fig. 8. 600-GHz SHHD and 300-GHz fundamental-heterodyne Fourier spectra and corresponding reconstructed images of a part of the 1951 USAF chart. (a) Intensity and (b) phase Fourier spectrum at 600 GHz; (c) Intensity and (d) phase reconstruction of (a) and (b); (e) Intensity and (f) phase Fourier spectrum at 300 GHz. (g) Intensity and (h) phase reconstruction of (e) and (f). The images of (c) and (d) are the same as those shown in Fig. 6(c) and (g).

## 4. Discussion and conclusion

Here, we discuss advantages and challenges brought about by our two innovations – the use of the $2^{nd}$-order sub-harmonic heterodyne detection in Fourier imaging at sub-THz frequencies, and the introduction of the dual-substrate-lens approach to heterodyne detection.

Sub-harmonic detection allows to work with cheaper and more powerful LO radiation sources, but it comes at the price of an intrinsically lower small-signal conversion efficiency compared with heterodyne mixing at the fundamental frequency. Thus it is difficult to achieve enough dynamic range for Fourier imaging, for which one wishes to be able to detect all Fourier components of the object which have passed through the lens. The introduction of the front-side wax/PTFE superstrate lens on the TeraFET detector over-compensates the lower conversion efficiency by tight focusing of the LO radiation and enlarging the effective antenna area of the detector. With the resultant dual-substrate lens approach, we have implemented a $2^{nd}$-order sub-harmonic detection scheme for Fourier imaging. In our validation measurements at 600/300 GHz, we have reached a 60-dB dynamic range using 56-$\mu$W of 600-GHz illuminating radiation and 600-$\mu$W of 300-GHz LO radiation. The dynamic range has been found to be sufficient to allow the detection of all available Fourier components, thus leading to a diffraction-limited spatial resolution better than 0.5 mm. These results are to be compared with our previous work in [15] with a 300-GHz fundamental-heterodyne Fourier imaging system, which also reached a 60-dB dynamic range, but required 1 mW of illuminating radiation at 300 GHz (of which 60% was sacrificed due to attenuation for standing-wave suppression; the LO power was 432 $\mu$W). Both the 600-GHz system and the 300-GHz one yield images with a resolution at the diffraction limit. This implies that the 600-GHz imaging system reaches a spatial resolution which is twice better than that of the 300-GHz system. It furthermore has a 5.6-dB larger dynamic range as calculated for the same power values of the illumination beam and the LO beam.

However, the challenge calling for improvement is the problem of the noise and the phase distortion brought about by standing waves in the imaging system. Possible solutions are the shortening of the coherence length of the radiation, e.g. with the help of specially designed scattering elements [35] in the beam path in front of the imaging system, or the addition of polarization filters such as magneto-optical Faraday isolators or related devices in the imaging system. Other options for the (numerical) suppression of standing-wave artifacts such as performing measurements at different distances or frequencies are also feasible, but would considerably enhance the measurement time.

There are other intriguing options to explore in the future. One relates to the sparsity of the Fourier spectrum. Dedicated inspection tasks may be performed with the recording of a minimal amount of pixels. Another option is the development of a comprehensive image reconstruction algorithm which includes aberration correction and allows for super-resolution. A third (and the most ambitious) one is the quest for phase retrieval with the help of deep-learning techniques, with the longer-term goal to replace heterodyne detection with mere power detection. In the case of success, one will not need a LO source anymore and can employ multi-pixel focal-plane arrays as power-detecting cameras, with all possible benefits such as real-time THz imaging and coverage of a large field of view. We have taken a first step in this general direction with a study on sub-THz inline holography at the Fresnel diffraction distance, where we have demonstrated successful phase retrieval from power-detection images using physics-informed supervised and unsupervised deep learning [36].

## Acknowledgment

This work was funded by the German Research Foundation (DFG) under Grant RO 770/48-1. Alvydas Lisauskas is thankful for the support received from the Foundation for Polish Science, grant IRA CENTERA. The authors would like to thank M. Zhang and D. Erni (Institute for General and Theoretical Electrical Engineering, University of Duisburg-Essen) for their help with the design of the dual substrate lens.

## References

1. G. Valušis, A. Lisauskas, H. Yuan, W. Knap, and H. G. Roskos, "Roadmap of terahertz imaging 2021," Sensors **21**, 4092 (2021).
2. F. Ellrich, M. Bauer, N. Schreiner, A. Keil, T. Pfeiffer, J. Klier, S. Weber, J. Jonuscheit, F. Friederich, and D. Molter, "Terahertz quality inspection for automotive and aviation industries," J. Infrared Millim. THz Waves **41**, 470–489 (2018).
3. M. Vandewal, E. Cristofani, A. B. W. Vleugels, F. Ospald, R. Beigang, S. Wohnsiedler, C. Matheis, J. Jonuscheit, J. P. Guillet, B. Recur, P. Mounaix, I. M. Hönninger, P. Venegas, I. Lopez, R. Martinez, and Y. Sternberg, "Structural health monitoring using a scanning THz system," in *38th International Conference on Infrared, Millimeter, and Terahertz Waves (IRMMW-THz),* (2013), pp. 1 –2.
4. D. Voß, W. Zouaghi, M. Jamshidifar, S. Boppel, C. McDonnell, J. R. P. Bain, N. Hempler, G. P. A. Malcolm, G. T. Maker, M. Bauer, A. Lisauskas, A. Rämer, S. A. Shevchenko, W. Heinrich, V. Krozer, and H. G. Roskos, "Imaging and spectroscopic sensing with low-repetition-rate terahertz pulses and GaN TeraFET detectors," J. Infrared Millim. THz Waves **39**, 262–272 (2018).
5. N. V. Petrov, M. S. Kulya, A. N. Tsypkin, V. G. Bespalov, and A. Gorodetsky, "Application of terahertz pulse time-domain holography for phase imaging," IEEE Trans. Terahertz Sci. Technol. **6**, 464–472 (2016).
6. B. Recur, L. Frederique, J. B. Perraud, J. P. Guillet, I. Manek-Honninger, P. Desbarats, and P. Mounaixa, "3D millimeter waves Tomosynthesis for the control of aeronautics materials," in *38th International Conference on Infrared, Millimeter, and Terahertz Waves (IRMMW-THz),* (2013), p. 1.
7. D. Simic, K. Guo, and P. Reynaert, "A 0.42 THz coherent TX-RX system achieving 10dBm EIRP and 27dB NF in 40nm CMOS for phase-contrast imaging," in *IEEE International Solid-State Circuits Conference (ISSCC),* (2021), pp. 318 –320.
8. S. M. H. Naghavi, S. Seyedabbaszadehesfahlani, A. C. F. Khoeini, and E. Afshari, "A 250 GHz autodyne FMCW radar in 55nm BiCMOS with micrometer range resolution," in *IEEE International Solid- State Circuits Conference (ISSCC),* (2021), pp. 320– 322.
9. W. L. Chan, M. L. Moravec, R. G. Baraniuk, and D. M. Mittleman, "Terahertz imaging with compressed sensing and phase retrieval," Opt. Lett. **33**, 974–976 (2008).
10. C. M. Watts, D. Shrekenhamer, J. Montoya, G. Lipworth, J. Hunt, T. Sleasman, S. Krishna, D. R. Smith, and W. J. Padilla, "Terahertz compressive imaging with metamaterial spatial light modulators," Nat. Photon. **8**, 605–609 (2014).

11. S. Venkatesh, H. S. X. Lu, and K. Sengupta, "A high-speed programmable and scalable terahertz holographic metasurface based on tiled CMOS chips," Nat. Electron. **3**, 785–793 (2020).
12. M. Humphreys, J. P. Grant, I. Escorcia-Carranza, C. Accarino, M. Kenney, Y. D. Shah, K. G. Rew, and D. R. S. Cumming, "Video-rate terahertz digital holographic imaging system," Opt. Express **26**, 25805–25813 (2018).
13. M. Locatelli, M. Ravaro, S. Bartalini, L. Consolino, R. Vitiello, Miriam S.and Cicchi, F. Pavone, and P. De Natale, "Real-time terahertz digital holography with a quantum cascade laser," Sci. Rep. **5**, 13566 (2015).
14. A. Siemion, L. Minkevicius, D. Jokubauskis, R. Ivaskeviciute-Povilauskiene, and G. Valusis, "Terahertz digital holography: Two- and four-step phase shifting technique in two plane image recording," AIP Adv. **11**, 105212 (2021).
15. H. Yuan, D. Voß, A. Lisauskas, D. Mundy, and H. G. Roskos, "3D Fourier imaging based on 2D heterodyne detection at THz frequencies," APL Photonics **4**, 106108 (2019).
16. H. Guerboukha, K. Nallappan, and M. Skorobogatiy, "Exploiting k-space/frequency duality toward real-time terahertz imaging," Optica **5**, 109–116 (2018).
17. J. P. Caumes, A. Younus, S. Salort, B. Chassagne, B. Recur, A. Ziéglé, A. Dautant, and E. Abraham, "Terahertz tomographic imaging of XVIIIth Dynasty Egyptian sealed pottery," Appl. Opt. **50**, 3604–3608 (2011).
18. N. George, *Fourier Optics* (Book published online, 2012). Available at http://www.hajim.rochester.edu/optics/sites/george/FO_BOOK.pdf.
19. F. Friederich, W. von Spiegel, M. Bauer, F. Meng, M. D. Thomson, S. Boppel, A. Lisauskas, B. Hils, V. Krozer, A. Keil, T. Loeffler, R. Henneberger, A. K. Huhn, G. Spickermann, P. H. Bolivar, and H. G. Roskos, "Thz active imaging systems with real-time capabilities," IEEE Trans. Terahertz Sci. Technol. **1**, 183–200 (2011).
20. A. Lisauskas, S. Boppel, M. Mundt, V. Krozer, and H. G. Roskos, "Subharmonic mixing with field-effect transistors: Theory and experiment at 639 GHz high above $f_T$," IEEE Sensors J. **13**, 124–132 (2013).
21. V. Giliberti, A. D. Gaspare, E. Giovine, S. Boppel, A. Lisauskas, H. G. Roskos, and M. Ortolani, "Heterodyne and subharmonic mixing at 0.6 THz in an AlGaAs/InGaAs/AlGaAs heterostructure field effect transistor," Appl. Phys. Lett. **103**, 093505 (2013).
22. H. Yuan, A. Lisauskas, M. Zhang, A. Rennings, D. Erni, and H. G. Roskos, "Dynamic-range enhancement of heterodyne THz imaging by the use of a soft paraffin-wax substrate lens on the detector," in *2019 Photonics & Electromagnetics Research Symposium - Fall (PIERS - Fall),* (2019), pp. 2607–2611.
23. S. Thomas, C. Bredendiek, and N. Pohl, "A sige-based 240-ghz fmcw radar system for high-resolution measurements," IEEE Trans. on Microw. Theory Tech. **67**, 4599–4609 (2019).
24. K. Ikamas, D. Čibiraitė, A. Lisauskas, M. Bauer, V. Krozer, and H. G. Roskos, "Broadband terahertz power detectors based on 90-nm silicon CMOS transistors with flat responsivity up to 2.2 THz," IEEE Electron Device Lett. **39**, 1413–1416 (2018).
25. D. Glaab, S. Boppel, A. Lisauskas, U. Pfeiffer, E. Oejefors, and H. G. Roskos, "Terahertz heterodyne detection with silicon field-effect transistors," Appl. Phys. Lett. **96**, 042106 (2010).
26. A. Lisauskas, M. Bauer, S. Boppel, M. Mundt, B. Khamaisi, E. Socher, R. Venckevicius, L. Minkevicius, I. Kasalynas, D. Seliuta, G. Valusis, V. Krozer, and H. G. Roskos, "Exploration of terahertz imaging with silicon MOSFETs," J. Infrared Millim. THz Waves **35**, 63–80 (2014).
27. H. Yuan, M. Zhang, Q. ul Islam, D. Erni, and H. G. Roskos, "A wax/PTFE substrate lens enables effective front-side illumination of TeraFET detectors," Manuscript in preparation.
28. P. Vizmuller, *RF Design Guide: Systems, Circuits, and Equations*, Artech House Antennas and Propagation Library (Artech House, 1995).
29. M. Bauer, A. Rämer, S. A. Chevtchenko, K. Y. Osipov, D. Čibiraitė, S. Pralgauskaitė, K. Ikamas, A. Lisauskas, W. Heinrich, V. Krozer, and H. G. Roskos, "A high-sensitivity algan/gan hemt terahertz detector with integrated broadband bow-tie antenna," IEEE Trans. on Terahertz Sci. Technol. **9**, 430–444 (2019).
30. D. Glaab, S. Boppel, A. Lisauskas, U. Pfeiffer, E. Öjefors, and H. G. Roskos, "Terahertz heterodyne detection with silicon field-effect transistors," Appl. Phys. Lett. **9**, 042106 (2010).
31. W. L. Chan, K. Charan, D. Takhar, K. F. Kelly, R. G. Baraniuk, and D. M. Mittleman, "A single-pixel terahertz imaging system based on compressed sensing," Appl. Phys. Lett. **93**, 121105 (2008).
32. Y. Zou, L. Zhang, C. Liu, B. Wang, Y. Hu, and Q. Chen, "Super-resolution reconstruction of infrared images based on a convolutional neural network with skip connections," Opt. Lasers Eng. **146**, 106717 (2021).
33. F. Wang, C. Wang, M. Chen, W. Gong, Y. Zhang, S. Han, and G. Situ, "Far-field super-resolution ghost imaging with a deep neural network constraint," Light. Sci. & Appl. **11**, 1 (2022).
34. H. Yuan, A. Lisauskas, M. Wan, J. T. Sheridan, and H. G. Roskos, "Resolution enhancement of THz imaging based on Fourier-space spectrum detection," in *Proc. SPIE, Terahertz, RF, Millimeter, and Submillimeter-Wave Technology and Applications XIII,* SPIE, ed. (2020), 1127918.
35. D. Li, D. P. Kelly, and J. T. Sheridan, "Speckle suppression by doubly scattering systems," Appl. Opt. **52**, 8617–8626 (2013).
36. M. Xiang, H. Yuan, L. Wang, K. Zhou, and H. G. Roskos, "Amplitude/phase retrieval for terahertz holography with supervised and unsupervised physics-informed deep learning," https://arxiv.org/abs/2212.06725 (2022).